\documentstyle[pre,multicol,aps]{revtex}

\newcommand \bq {\begin{eqnarray}}
\newcommand \be {\begin{equation}}
\newcommand \eq {\end{eqnarray}}
\newcommand \ee {\end{equation}}

\newcommand \ran {\rangle}

\newcommand \lan {\langle}

\newcommand \th {\theta}
\newcommand \r {\rho}

\begin{document}
\draft

\setcounter{page}{1}
\title{ About optimal measurements in quantum
hypothesizes testing.}

\author{A.E. Allahverdyan, D.B. Saakian }
\address{ Yerevan Physics Institute \\
Alikhanian Brothers St.2, Yerevan 375036, Armenia}
\maketitle
\begin{abstract}
We consider the problem of a state determination 
for a two-level quantum system which can be in one of two nonorthogonal
mixed states. It is proved that for the two independent identical  
systems the optimal combined measurement (which considers the pair 
as one system) cannot be less optimal than the corresponding
sequential one (local measurements, accompanying by transfer of classical information).
The case of equality is achieved only when the mixed states have the same 
eigenvalues or the same eigenvectors. Further, we consider a case
then the two systems are entangled: measurement of one system
induces a reduction of the another one's state. The conclusion about optimal
character of  combined measurement takes place again, and conditions where the
above-mentioned methods coincide are derived.
\end{abstract}

\vspace{8mm}
\begin{multicols}{2} 
A  determination of an unknown quantum state is one of the most
important problems in  modern quantum theory
\cite{peres}\cite{massar}\cite{brody}\cite{japan}\cite{helstrom}.
It is usually assumed that there are unlimited number of
identical and independent systems in the same state (replicas).
Then an estimation of the unknown state with arbitrary precision is 
possible. However, the well-known theorem \cite{zurek-barnum} forbids 
cloning of any unknown state; thus for generation of sufficiently 
large number of replicas the exact control of the corresponding 
equipment should be achieved. 
In any case it is reasonable to consider 
a finite number of different replicas, i.e., some error in 
determination of the quantum state is unavoidable. On the 
other hand, different replicas can be correlated. Particularly,
this fact can be viewed against exhaustive interpretation of
probabilities (quantum or classical) in spirit of the law of
large numbers \cite{jaynes}. 

The particular case  of an unknown state determination is
quantum hypothesizes testing: It is assumed that the system can
be only in one of some non-orthogonal states. 
The similar
situation occurs also in quantum information theory where
classical information is transmitted by means of some
non-orthogonal states \cite{helstrom}\cite{caves}\cite{allah}.
Generally speaking, nonorthogonality can arise due to noises in the
transmitter or energy loss in the channel. Also it can be 
connected with the construction of the transmitter,
which generates coherent states for example. It is not necessary
to consider non-orthogonality as some hindrance only: In some cases
(such as quantum cryptography or some noisy channels) non-orthogonal
states can be more useful.
We shall consider the case of {\it bayesian} hypothesizes testing where some {\it a priori} 
probabilities are assumed for the each hypothesis. Particularly, information from 
preceding experiments can be accumulated at these probabilities.
Let a two-level quantum system (qubit) $A$ can be in two different
non-orthogonal mixed states: $\rho _1$ and $\rho _2$ with equal a priori probabilities.
A measurement is taken, which is described by operators $\Pi _1$ and $\Pi _2$ 
($\Pi _1+\Pi _2$=1), for distinguishing between them. The probability of
registration a state $\rho _k$ if the initial state was $\rho _i$ is $p(k/i)={\rm tr}(\rho _i\Pi _k)$.
The optimal measurement must minimize the mean probability of error $P_e$,
\bq
\label{3}
&&P_e=\frac{1}{2}(p(2/1)+ p(1/2))=\frac{1}{2}(1+{\rm tr}(\r \Pi _2)),\nonumber \\ 
&&\r =\r _1-\r _2
\eq
We see that ${\rm tr}(\r \Pi _2)=\sum_i(\r \Pi _2)_{(i)}$ 
should be minimized (where $(\r \Pi _2)_{(i)}$ is the 
corresponding eigenvalue). 
Thus the resulting formulas are \cite{helstrom}
\bq
\label{5}
& & \Pi _2 ^{({\rm opt})}=\sum_i\th (-\r _{(i)})|\r _{(i)}\ran \lan \r _{(i)}|, \nonumber \\
& & P _e ^{({\rm opt})}(\r _1, \r_2)=\frac{1}{2}(1+\sum_i\th (-\r _{(i)})\r _{(i)}).
\eq
The optimal measurement allows to rederive  a  priori probabilities. If $\bar{p}_s(k)$ is the
probability of the initial realization of $\rho _k$ if $\rho _s$ was recorded, then according
to the famous Bayes-Laplace formula we have
\be
\label{n6}
\bar{p}_s(k)=\frac{1}{2}\frac{p(s/k)}{p(s)},
\ee
where $p(s)=(p(s/1)+p(s/2))/2$ are the total output probabilities. Thus, if the result
of the measurement is $s$ as {\it post priori} probabilities can be chosen 
$\bar{p}_s(1)$ and $\bar{p}_s(2)$.

Now let us assume that there is exactly the same system $B$ which is 
independent from the previous one. If the method of  measurement allows to
consider the systems $A$, $B$ as  one single system 
(combined or global measurements) we have the case of distinguishing between $\r _1\otimes\r _1$
and $\r _2\otimes\r _2$; so the corresponding minimal mean error is 
\be
\label{n7}
P_{e,g}=P _e ^{({\rm opt})}(\r _1\otimes\r _1 ,\r _2\otimes\r _2). 
\ee
However, in practice combined measurements cannot be realized in many cases
(for example, if $A$ and $B$ is separated by a sufficiently
large distance). In such cases an other scheme  can be proposed:
Sequential (or local, accompanying by transfer of classical information) measurements. 
There are two different observers for each replica, and the following steps are realized: 
To measure the system $A$, to obtain the corresponding post priori probabilities, after this
to send this (classical) information to a $B$-observer, which can use them as the  a priori
probabilities for his measurement. The mean error in this scheme must depend on an outcome
of the first measurement, thus the total mean error probability can be defined by averaging with the
output probabilities $p(s)$ of the first measurement.
The resulting formula is
\bq
\label{n8}
&&P_{e,l}=\frac{1}{2}+\frac{1}{2}\sum_{s=1}^2\, 
p(s/1)\sum_k\th (-\r _{(k)}(s))\r _{(k)}(s), \nonumber \\
&&\r (s)=\r _1-\lambda (s)\r _2, \, \lambda (s)=\frac{p(s/2)}{p(s/1)}.
\eq
First, the physical difference between combined and sequential measurements has been pointed out
by Peres and Wootters \cite{peres}. Here the case of three linearly-dependent pure-state
hypothesizes was investigated, and showed, by numerical methods, that combined measurements 
are more optimal. For a more practically important case of two linearly-independent pure-states 
hypothesizes it was shown \cite{brody}\cite{japan} that these methods  coincide 
(in the sense of the mean probability of error \cite{brody}, as well as in the 
sense of mutual information \cite{japan}).
For the same case, but  using a quantity which is neither the
mutual information nor the mean probability of error, the problem has been considered 
in Ref. \cite{massar}. It was claimed again that combined measurements are more optimal.
The optimal character of combined measurements has been stressed very recently also
\cite{ben2}. In the present paper we consider mixed states and mean probability of error as 
the measure of distinguishability; the problem in more 
general settings remains open.

For calculating $P_{e,l}$, $P_{e,g}$ it is convenient to choose the base where the matrix 
$\r =\r _1-\r _2$ is the diagonal one. Thus we have ($k=1,2,
\, \langle 0|\bar{0} \rangle =\delta _{\bar{0} 0}$):
\be
\label{n9}
\r _k=x_k|0\rangle \langle 0| +(1-x_k)|\bar{0}\rangle \langle \bar{0}| +z|0\rangle \langle \bar{0}|
+z^{*}|\bar{0}\rangle \langle 0|.
\ee
After calculations (which are not reproduced here) we get
\be
\label{n10}
P_{e,g}=\frac{1}{2}(1-|x_1-x_2|(|x_1+x_2-1|+\sqrt{1+4|z|^2})),
\ee
\bq
\label{n11}
&&P_{e,l}=\frac{1}{2}-\frac{|x_1-x_2|}{4} \nonumber \\
&&(\sqrt{1+4[|z|^2+|x_1+x_2-1|^2+|x_1+x_2-1|]} \nonumber \\
&&+\sqrt{1+4[|z|^2+|x_1+x_2-1|^2-|x_1+x_2-1|]_+} ),
\eq
where $[x]_+=(x+|x|)/2$. In the general case
it can be proved exactly that $P_{e,g}\leq P_{e,l}$, i.e., 
the combined measurement is more optimal.
The equality is achieved in the following cases: 1) $\r _1 =\r _2$ 
(it is the trivial 
and non-interesting case where the state is known exactly);
2) $x_1+x_2=1$, when the matrices $\r _1$, $\r _2$ have the same eigenvalues; 
3) $z=0$, when $\r _1$ and 
$\r _2$ commute (i.e., the corresponding eigenvalues can be obtained simultaneously). 
All these cases can be jointed saying that $P_{e,g}=P_{e,l}$ holds if the matrices have the same 
eigenvalues or the same eigenvectors. 
It seems to us that these results are more general ones but we have not
succeeded in proving it.
Particularly, the statement holds for two pure-state hypothesizes 
\cite{brody}\cite{japan}.

We have considered the case of two uncorrelated replicas of the same system. 
Now let us consider the case 
where replicas $A$, $B$ are entangled. As it was stated above, this fact can be connected with 
the corresponding equipment which prepares  replicas. 
The similar statement of problem occurs also in the broadcast 
quantum-classical communication \cite{allah}, and more generally in cases when
a many-particle quantum system is measured.

Let us assume that the pairs of replicas can be in the following pure states:
\bq
\label{n12}
&&|\psi _k\rangle =\alpha _k|0\rangle _{A}|0 \rangle _{B} +
\gamma _k|\bar{0}\rangle _{A}| \bar{0} \rangle_{B}  \nonumber \\
&&+\beta _k|0\rangle _{A}  |\bar{0}\rangle _{B}
+\tilde{\beta _k }|\bar{0}\rangle _{A} |0 \rangle _{B}, \, k=1,2.
\eq
The following restriction is added
\be
\label{n13}
{\rm tr}_{A(B)}|\psi _k\rangle \langle \psi _k|=\r _k,
\ee
in other words the replicas are equivalent  although they are entangled
(i.e., cannot be moved to a product-state using local transformations {\it only}).
For this condition it is sufficient to choose $\tilde{\beta _k }=\beta _k$. The base 
$|0\rangle $, $|\bar{0}\rangle $
is again the eigenbase of the matrix $\r =\r _1-\r _2$: It holds
\be
\label{dop1}
\alpha _{1}\beta^{*} _{1}+\beta _{1}\gamma^{*} _{1}
=\alpha _{2}\beta ^{*} _{2}+\beta _{2}\gamma^{*} _{2}.
\ee
It should be noted that we choose the {\it local} base, therefore the entanglement conserves.

The method of combined measurement is unchanged: the states $|\psi _1\rangle $, 
$|\psi _2\rangle $ must be distinguished with equal a priori probabilities; thus the 
corresponding mean probability of error is 
\be
\label{n14}
P_{e,g}=\frac{1}{2}(1-\sqrt{1-|\tau |^2}), \, \tau =\langle\psi _2|  \psi _1 \rangle
\ee
It is obvious, that after a measurement of $A$ a reduction of the 
general system's state occurs.
We shall describe this reduction as the standard von Neumann's one. However, 
it is extremely important
to recognize that the reduction of this type is not realized for {\it arbitrary} quantum measurement.
Furthermore, for its realization a special control to system-apparatus interaction must 
be ensured (for recent discussion see \cite{ozawa}).

After the optimal measurement of $A$ the total wave-function is changed as
\be
\label{n15}
|\psi _k\rangle \to |\psi _k^{(s)}\rangle =
\frac{ \Pi _s ^{({\rm opt})} |\psi _k\rangle }
{\sqrt{\langle \psi _k| \Pi _s ^{({\rm opt})} |\psi _k\rangle }}, \, k,s=1,2,
\ee
(where $ \Pi _s ^{({\rm opt})}$ acts only on $A$-coordinates). Thus the $B$-observer
"sees" the states
\be
\label{n16}
\r _k^{(s)}={\rm tr}_{A}(|\psi _k^{(s)}\rangle \langle \psi _k^{(s)}|).
\ee
In the first place, the optimal measurement is performed on the system $A$. 
Thus,  a priori probabilities can be rederived to post priori ones and sent to $B$-observer, 
who should determine the
actual state of  his sub-system. 
(The similar separation of information to quantum one, realizing through the 
reduction, and classical one takes place, for example, in quantum teleportation \cite{ben}.)
As it is usual, at the final stage the result should be averaged by $p(s)$. Thus it holds
\bq
\label{n17}
&&P_{e,l}=\frac{1}{2}+\frac{1}{2}\sum_{s=1}^2\, 
p(s/1)\sum_k\th (-\r ^{(s)}_{(k)}(s))\r^{(s)} _{(k)}(s), \nonumber \\
&&\r^{(s)}(s)=\r^{(s)} _1-\lambda (s)\r^{(s)} _2, \, \lambda (s)=\frac{p(s/2)}{p(s/1)}.
\eq
With (\ref{n12}) we get 
\bq
\label{n18}
&&P_{e,l}=\frac{1}{2}-\frac{1}{4}(\sqrt{(x_1-x_2)^2 +4|\alpha _1\beta _2-\alpha _2\beta _1|^2}
\nonumber \\
&&+\sqrt{(x_1-x_2)^2 +4|\gamma _1\beta _2-\gamma _2\beta _1|^2 }), \nonumber \\
&& x_1=|\alpha _1|^2+|\beta _1|^2, \, x_2=|\alpha _2|^2+|\beta _2|^2.
\eq
Our central problems are to compare $P_{e,l}$ and $P_{e,g}$, and to derive the conditions
when these quantities coincide.
\bq
\label{n19}
&&(1-2P_{e,l})^2-(1-2P_{e,g})^2\nonumber \\
&&=u_1(u_2-1-\bar{\tau})+u_2(u_1-1+\bar{\tau})+|\tau |^2-\bar{\tau }^2 \nonumber \\
&&+2\sqrt{u_1u_2(1-\bar{\tau }-u_1 )(1+\bar{\tau }-u_2) } ,
\eq
where we have introduced the following notations:
\bq
\label{n20}
&&2u_{1,2}=x_1+x_2\mp 2|\alpha _1\alpha^{*} _2+\beta _1\beta^{*} _2|, \nonumber \\
&&\bar{\tau}=
|\alpha _1\alpha^{*} _2+\beta _1\beta^{*} _2|+|\gamma _1\gamma^{*} _2+\beta _1\beta^{*} _2|.
\eq
If $\bar{\tau}=|\tau |$
when the right-hand side of Eq. (\ref{n19}) is always non-positive, 
it becomes exactly negative if $\bar{\tau}\not =|\tau |$
because now $\bar{\tau}>|\tau |$. Thus the combined method
stills more optimal one, and the methods are equitable if
\be
\label{n21}
\bar{\tau}=|\tau |, \, u_1(1+|\tau|)=u_2(1-|\tau|).
\ee
These conditions are satisfied automatically if $|\psi _1\rangle$, $|\psi _2\rangle$
can be presented as direct products. Conditions (\ref{n21}) can be written in a more
subtle way:
\be
\label{n22}
{\rm arg} (\langle \psi _2|\Pi _1 ^{({\rm opt})}|\psi _1\rangle) =
{\rm arg}(\langle \psi _2|\Pi _2 ^{({\rm opt})}|\psi _1\rangle ),
\ee
\bq
\label{n23}
&&2|\langle \psi _2|\Pi _s ^{({\rm opt})}|\psi _1\rangle |=|\langle \psi _2|\psi _1\rangle |
(\langle \psi _1|\Pi _s ^{({\rm opt})}|\psi _1\rangle \nonumber \\
&&+\langle \psi _2|\Pi _s ^{({\rm opt})}|\psi _2\rangle )
\eq
(the conditions with $s=1,2$ are equivalent). 
We see  that an entanglement between replicas can make
the combined method more optimal than the sequential one. 
The detailed investigation of Eqs. (\ref{n22})(\ref{n23}) will be given elsewhere.
Let us  mention only some special cases
(among many others) where the conditions (\ref{n22})(\ref{n23}) are satisfied
(it is assumed for simplicity that all parameters are real, and $|\psi _k\rangle $ 
are correctly normalized):
 1) $\alpha _1=\gamma _2$,  $\alpha _2=\gamma _1$, $\beta _1=\beta _2$; 
 2) $\alpha _1=-\gamma _1$, $\alpha _2=-\gamma _2$;
 3) $\alpha _1=\gamma _1$, $\alpha _2=\gamma _2$, $\alpha _1\beta _1=\alpha _2\beta _2$.

We have proved that for a binary-state unknown system the combined measurement is more
optimal than the sequential one, excepting two special cases where the corresponding mixed
states have the same eigenvectors or the same eigenvalues. Also we prove that this conclusion
about the combined method takes place again if the replicas of the unknown system are entangled
(excepting the special cases described by Eqs. (\ref{n22})(\ref{n23})).

This work was inspirited by the brilliant paper of E.T. Jaynes \cite{jaynes}, where a very elegant
and interesting discussion about quantum statistical description can be found.



\end{multicols} 

\end{document}